\documentclass[journal]{IEEEtran}

\usepackage[T1]{fontenc}
\usepackage{amsmath,amssymb,bm}
\usepackage{booktabs,multirow,tabularx,array}
\usepackage{algorithm}
\usepackage{algpseudocode}
\usepackage{tikz}
\usetikzlibrary{arrows.meta,positioning,shapes.geometric,fit,backgrounds}
\usepackage{pgfplots}
\usepackage{pgfplotstable}
\usepgfplotslibrary{groupplots}
\pgfplotsset{compat=1.18}
\usepackage[caption=false,font=footnotesize]{subfig}
\usepackage{cite}
\usepackage{xurl}
\usepackage[hidelinks]{hyperref}

\newcommand{\FGRExactRecall}{100.00\%}
\newcommand{\FGRExactPerfectQueries}{180}
\newcommand{\FGRCertRecall}{99.91\%}
\newcommand{\FGRCertPrecision}{100.00\%}
\newcommand{\FGRCertBound}{99.41\%}
\newcommand{\FGRCertGap}{0.50~pp}
\newcommand{\FGRCertMinBound}{95.26\%}

\newcommand{\FGRCertViolationCount}{0}
\newcommand{\FGRCertBelowTargetCount}{0}
\newcommand{\BoundAuditComparisons}{6,080}
\newcommand{\BoundAuditMinSlack}{$3.56\times 10^{-12}$}
\newcommand{\FGRExactMedianMs}{7.27}
\newcommand{\FGRCertMedianMs}{7.24}
\newcommand{\FGRCertPNinetyFiveMs}{12.39}
\newcommand{\SpatialMedianMs}{1.24}
\newcommand{\SpatialPNinetyFiveMs}{1.99}
\newcommand{\LatencyRatio}{5.85}
\newcommand{\PNinetyFiveLatencyRatio}{6.22}
\newcommand{\FGRExaminedSaving}{1.1}

\newcommand{\PeriodicDaySixtyRecall}{77.89\%}
\newcommand{\PeriodicDaySixtyPrecision}{82.44\%}
\newcommand{\MedianReplayRate}{903}
\newcommand{\MaxReplayRate}{1025}
\newcommand{\FGRSerializedMB}{5.39}
\newcommand{\SpatialSerializedMB}{5.18}
\newcommand{\FrontierFGRZeroRecall}{77.88\%}
\newcommand{\FrontierFGRZeroCert}{55.63\%}
\newcommand{\FrontierFGRNinetyFiveRecall}{100.00\%}
\newcommand{\FrontierFGRNinetyFiveCert}{99.71\%}
\newcommand{\FrontierFGRMinLatencyMs}{5.56}
\newcommand{\FrontierFGRMaxLatencyMs}{6.85}
\newcommand{\FrontierIVFMaxRecall}{100.00\%}
\newcommand{\FrontierGraphMaxRecall}{37.46\%}
\newcommand{\StatsDiffLIST}{+0.143}
\newcommand{\StatsDiffIVF}{+0.145}
\newcommand{\StatsDiffGraph}{+0.673}
\newcommand{\StatsDiffExact}{-0.00092}
\newcommand{\StatsMinHolmP}{0.5625}
\newcommand{\RadiusOneKRecall}{100.00\%}
\newcommand{\RadiusOneKCert}{100.00\%}
\newcommand{\RadiusFiveKLatencyMs}{8.45}
\newcommand{\RadiusFiveKRecall}{99.90\%}
\newcommand{\RadiusFiveKCert}{99.00\%}
\newcommand{\ThresholdLowAnswers}{148.3}
\newcommand{\ThresholdHighAnswers}{29.6}
\newcommand{\ThresholdLowLatencyMs}{10.29}
\newcommand{\ThresholdHighLatencyMs}{5.44}
\newcommand{\ThresholdLowRecall}{99.83\%}
\newcommand{\ThresholdHighRecall}{100.00\%}
\newcommand{\ThresholdLowCert}{98.97\%}
\newcommand{\ThresholdHighCert}{99.81\%}
\newcommand{\ReplayFiftyMs}{55.4}
\newcommand{\ReplayOneHundredMs}{108.2}
\newcommand{\ReplayTwoFiftyMs}{285.0}
\newcommand{\ReplayFiveHundredMs}{576.5}
\newcommand{\ReplaySevenFortyMs}{721.6}
\newcommand{\RebuildMinMs}{384.6}
\newcommand{\RebuildMaxMs}{509.8}
\newcommand{\ReplayLastFasterEvents}{250}
\newcommand{\ReplayFirstSlowerEvents}{500}
\newcommand{\StaleFGRMinRecall}{100.00\%}
\newcommand{\PeriodicDayThirtyFiveRecall}{98.95\%}
\newcommand{\PeriodicDayFortyRecall}{95.52\%}
\newcommand{\PeriodicDayFortyFiveRecall}{81.40\%}
\newcommand{\PeriodicDayFiftyRecall}{95.81\%}
\newcommand{\DriftDayFortyJSD}{0.09}
\newcommand{\DriftDayFortyFiveJSD}{0.55}
\newcommand{\DriftRebuildMs}{298.3}
\newcommand{\DriftDayFortyFivePNinetyFiveMs}{7.43}
\newcommand{\DriftDayFiftyPNinetyFiveMs}{5.85}
\newcommand{\DriftDayFortyFiveCert}{99.43\%}
\newcommand{\DriftDayFiftyCert}{98.09\%}
\newcommand{\SizeFGRFiveHundredMB}{3.88}
\newcommand{\SizeFGRTwoThousandMB}{5.01}
\newcommand{\SizeSpatialFiveHundredMB}{3.82}
\newcommand{\SizeSpatialTwoThousandMB}{4.88}
\newcommand{\SelectivityFGREmptyMs}{5.93}
\newcommand{\SelectivityFGRHighMs}{9.23}
\newcommand{\SelectivitySpatialMinMs}{1.15}
\newcommand{\SelectivitySpatialMaxMs}{1.39}
\newcommand{\SelectivityFGRLowRecall}{100.00\%}
\newcommand{\SelectivityFGRHighRecall}{99.67\%}
\newcommand{\EnglishRawRecall}{60.51\%}
\newcommand{\FrenchRawRecall}{56.82\%}
\newcommand{\FrenchLexiconRecall}{61.36\%}
\newcommand{\SpanishRawRecall}{56.53\%}
\newcommand{\SpanishLexiconRecall}{61.36\%}
\newcommand{\GermanRawRecall}{57.10\%}
\newcommand{\GermanLexiconRecall}{60.51\%}
\newcommand{\SynonymRawRecall}{55.97\%}
\newcommand{\SynonymLexiconRecall}{60.51\%}
\newcommand{\FullAblationRecall}{99.68\%}
\newcommand{\FullAblationCert}{99.20\%}
\newcommand{\FullAblationLatencyMs}{8.17}
\newcommand{\NoGraphAblationRecall}{99.68\%}
\newcommand{\NoGraphAblationCert}{99.20\%}
\newcommand{\NoGraphAblationLatencyMs}{6.93}
\newcommand{\NoSemanticAblationLatencyMs}{7.93}
\newcommand{\OneBlockAblationLatencyMs}{8.18}
\newcommand{\StaleNoDeltaRecall}{59.38\%}
\newcommand{\PrimaryRuns}{1800}
\newcommand{\QueriesPerMethod}{180}

\newcommand{\PilotRawObjects}{2,500}
\newcommand{\PilotBaseObjects}{2,000}

\newcommand{\PilotEvents}{740}
\newcommand{\UniquePrimaryQueries}{180}

\newcommand{\PilotCPU}{AMD EPYC 9V74 80-Core Processor}
\newcommand{\PilotLogicalCPUs}{9}
\newcommand{\PilotKernel}{6.18.35}
\newcommand{\PilotPython}{3.12.14}
\newcommand{\PilotNumpy}{2.3.5}
\newcommand{\PilotPandas}{2.2.3}
\newcommand{\PilotSklearn}{1.8.0}
\newcommand{\PilotScipy}{1.17.0}
\newcommand{\PilotPeakRSSKiB}{283,452}

\title{A Functional Pilot for Certified Freshness-Aware Semantic--Spatial Range Retrieval}

\author{Taimoor Ahmed \thanks{Superior University Lahore, Pakistan}}

\begin{document}

\maketitle

\begin{abstract}
Geographic applications need every object inside a radius that satisfies a semantic threshold, yet embedding indexes return approximate top-ranked lists and may omit qualifying records silently. We present FRESH-GEORANGE, a semantic-spatial range design that separates source-watermark freshness from optional record age. Geographic cells and semantic microblocks provide admissible pruning bounds; a graph proposes verification order but supplies no correctness evidence. Exact mode scans every nonprunable block and the delta overlay. Certified mode may stop early and reports a deterministic query-specific recall lower bound from verified answers and unresolved records. A reproducible CPU pilot uses \PilotRawObjects{} real OpenFlights airport records, a \PilotBaseObjects{}-record base, and \PilotEvents{} simulated insert, delete, and text-revision events; it evaluates \UniquePrimaryQueries{} unique queries over five seeds. Exact mode achieved \FGRExactRecall{} set recall on every query. The 95-percent mode achieved \FGRCertRecall{} empirical mean recall with a \FGRCertBound{} reported mean certificate and no observed bound violation. However, its \FGRCertMedianMs{} ms median latency was \LatencyRatio{} times the \SpatialMedianMs{} ms spatial-first exact baseline, and full-history delta replay became slower than rebuilding at larger batches. The prototype therefore validates the completeness mechanism, not performance superiority or production freshness. Submission-scale evaluation requires real map diffs, official recent baselines, and truly incremental versioned maintenance.

\end{abstract}

\begin{IEEEkeywords}
semantic--spatial retrieval, range reporting, freshness watermark, certified recall, dynamic index, exact verification
\end{IEEEkeywords}

\section{Introduction}
\label{sec:introduction}
Location-aware search increasingly depends on meanings beyond literal keyword overlap. A request for ``emergency treatment for a severe allergic reaction'' should retrieve a nearby facility advertising ``anaphylaxis services.'' Learned systems support such matches by embedding queries and points of interest (POIs). LIST couples learned routing with embedding relevance~\cite{yin2025list,a4}, while Mesh integrates geographic range constraints with high-dimensional approximate nearest-neighbor (ANN) search~\cite{song2025mesh}. Both target a ranked top-$k$ list. Operational tasks such as emergency-resource discovery and regulatory auditing instead need threshold enumeration---every active object inside a stated radius whose semantic similarity reaches a threshold. The caller must know whether the returned list is exhaustive, rather than merely well ranked; an incomplete top-$k$ list cannot safely represent that set~\cite{a92}.

For top-$k$, ANN may stop after collecting enough competitive candidates. Under a threshold, cardinality is unknown and every skipped partition could hide an answer. Exact rescoring removes false positives but cannot recover unseen positives; it certifies precision, not completeness. A valid certificate must therefore account for partitions that heuristic traversal never reaches, not infer completeness from the retrieved candidates. Range-complete foundations include Boolean radius engines~\cite{lee2015mainmemoryspatialkeyword}, WISK's learned partitions for exact region-and-keyword processing~\cite{sheng2023wisk}, and geographic-radius enumeration under a Jaccard threshold~\cite{tampakis2021spatialkeywordrange}. Word2Vec has also been used for continuous spatial range processing~\cite{oh2018word2vecrange,a4,a6}. These interfaces do not jointly define dense-semantic threshold enumeration, versioned visibility, and a query-specific completeness certificate.

Mutability creates a second gap. POI catalogs receive insertions, deletions, and text revisions continually, so an answer may be correct for an old index but wrong for the promised version. Per-edit rebuilding is costly; periodic rebuilding hides new resources and can resurrect deleted ones between rebuilds. Dynamic spatial--keyword systems maintain queries as objects change~\cite{mahmood2018fast,dong2021continuoustopk,salgado2018continuousrange}, while vector systems support streaming or incremental maintenance~\cite{xu2023spfresh,gong2025vstream,mohoney2025quake}. LIST is not purely static: it supports cluster-assigned insertion, deletion, and index-only retraining after sustained inserts or distribution shift~\cite{yin2025list,a7}. It nevertheless remains top-$k$ and supplies no radius-plus-threshold completeness contract. We therefore distinguish an update-visibility watermark, a correctness promise about reflected commits, from optional record-age semantics, an application predicate over old observations. A recent observation can be absent behind a stale watermark, while an old observation can be fully visible; treating both as one ``freshness'' value obscures correctness.

FRESH-GEORANGE addresses this conjunction. Its query specifies a location, radius, semantic vector, similarity threshold, and maximum visibility lag. The index combines geographic cells, semantic microblocks with admissible cosine bounds, an ANN graph used only to order work, and a latest-write delta overlay for inserts, replacements, and tombstones. A watermark identifies the visible snapshot. Exact mode visits every non-prunable spatially relevant block and scans the entire visible delta. Certified mode may stop earlier, but conservatively counts unresolved physical records and derives a deterministic per-query recall lower bound from that count and the verified positives. Tombstones and replacements suppress superseded base versions, and the delta and watermark are published atomically. Approximation therefore changes cost and certificate tightness, not the truth of the bound.

The closest adjacent field is range-filtered ANN (RFANN), which restricts dense-vector top-$k$ search by an interval on one ordered attribute. SeRF and iRangeGraph optimize static filters~\cite{zuo2024serf,xu2024irangegraph,a9}; Dynamic RFANNS, DIGRA, RangePQ, and WoW add dynamic or incremental maintenance~\cite{peng2025dynamicrfann,jiang2025digra,zhang2025rangepq,wang2025wow}. Yet an interval is not a two-dimensional geodesic disk, and approximate top-$k$ is not threshold enumeration. Mapping a disk to longitude intervals or a space-filling key introduces different selectivity and boundary behavior. Vector-radius retrieval approximately reports all vectors within an embedding distance but lacks a geographic predicate and mutable snapshot~\cite{manohar2025rangeretrieval}. ConANN provides a distribution-free conformal risk guarantee for ANN~\cite{horchidan2025conann}; our target is instead deterministic, bound-based accounting for one declared snapshot. We claim novelty only for this conjunction, not for any ingredient alone.

 We refer to this setting as semantic–spatial range retrieval, where the objective is not to return a fixed number of highly ranked objects, but to enumerate the complete set of objects satisfying both a geographic range constraint and a semantic-similarity threshold~\cite{manohar2025rangeretrieval}. This distinction is important because the number of qualifying objects is unknown before execution: a query may have no matches, a few matches, or hundreds of matches within the same radius. Consequently, conventional approximate nearest-neighbor search can miss qualifying objects that were never examined, even when all retrieved candidates are subsequently verified exactly~\cite{a6,a10,a11,a12}. We therefore use completeness to mean that every object satisfying the declared predicates is represented in the returned set. Dynamic data introduce an additional dimension, because completeness must be defined relative to a particular visible database state. We call the latest committed source position reflected by the index the visibility watermark, while record age describes the age of an individual stored observation; the two notions are intentionally separate. Finally, a recall certificate denotes a deterministic, query-specific lower bound on the fraction of qualifying objects that have been returned when execution terminates before exhaustive verification.

In our CPU pilot, exact mode attains \FGRExactRecall{} mean recall and certified mode attains \FGRCertRecall{} with a \FGRCertBound{} mean lower bound; however, certified retrieval costs \LatencyRatio$\times$ the median latency of spatial-first exact scanning. This negative result supports feasibility of the guarantee, not deployment superiority.

This paper makes three contributions:
\begin{itemize}
    \item We formalize \emph{fresh semantic--spatial range enumeration}: all visible POIs inside a geodesic radius and above a dense-similarity threshold, under an explicit update-watermark contract that distinguishes visibility lag from record age.
    \item We design a dual spatial--semantic index with a latest-write delta overlay, atomic watermark publication, and safe geometric and cosine bounds. Its ANN component only prioritizes verification; exact mode exhausts every non-prunable block, while certified mode returns a sound query-specific recall lower bound after exact delta scanning.
    \item We provide a reproducible CPU pilot that evaluates set recall and precision, certificate validity and tightness, tail latency, complete-history replay cost, staleness, scripted drift instrumentation, selectivity, and ablations under deterministic insertion, deletion, and text-revision streams, and we state the evidence still required for a submission-scale claim.
\end{itemize}

The remainder of the paper is organized as follows. Section~\ref{sec:related-work} positions the problem against spatial--textual, filtered-vector, and dynamic indexes. Section~\ref{sec:method} formulates the query and presents the index, update contract, and certification argument. Section~\ref{sec:experiments} describes the reproducible pilot and reports its results and limitations. Section~\ref{sec:conclusion} concludes.

\section{Related Work}
\label{sec:related-work}

\textbf{Spatial--textual retrieval.}
Classical structures combine a spatial hierarchy with an inverted text index. The IR-tree is a representative design for geographic document search~\cite{li2011irtree}; later main-memory processing supports point-plus-radius Boolean queries with perfect result recall~\cite{lee2015mainmemoryspatialkeyword}. WISK learns workload-aware spatial partitions while preserving exact keyword-query processing and buffers insertions before localized retraining~\cite{sheng2023wisk}. Tampakis et al. are especially close because their filter--refine index returns every object inside a radius whose keyword-set Jaccard similarity exceeds a threshold~\cite{tampakis2021spatialkeywordrange}. FRESH-GEORANGE changes the semantic predicate from a sparse token-set measure to dense cosine similarity and couples it to a mutable-snapshot certificate; it does not treat threshold range completeness itself as new.

Embedding-aware spatial retrieval has mainly evolved around ranking. The $S^2R$-tree uses semantic pivots for spatial keyword search~\cite{chen2020s2rtree}, DrW develops neural relevance modeling~\cite{liu2023drw}, and LIST jointly learns object/query representations, cluster routing, and spatial relevance for embedding-based top-$k$ spatial keyword queries~\cite{yin2025list}. LIST explicitly supports insertion, deletion, and index-only retraining; calling it static would erase a material capability. What it does not offer is all-object output under both a geographic radius and a similarity threshold, nor a per-query lower bound that accounts for work not visited. Mesh directly combines a true spatial range with vector ANN~\cite{song2025mesh}, making it the closest integrated vector baseline, but still returns approximate top-$k$ results. Oh et al. maintain Word2Vec-based range results over moving objects~\cite{oh2018word2vecrange}; FAST, Salgado et al., and Dong et al. likewise establish that continuous and dynamic spatial--keyword processing is mature~\cite{mahmood2018fast,salgado2018continuousrange,dong2021continuoustopk}. Their motion, Boolean, or top-$k$ semantics differ from versioned POI create/delete/text-replace operations with a declared visibility watermark.

\begin{table*}[t]
\centering
\caption{Closest-work landscape. ``All'' means threshold/range enumeration rather than a fixed top-$k$ list; guarantees are characterized by what the cited interface establishes. FRESH-GEORANGE's proposed novelty is the conjunction in the last row.}
\label{tab:closest-work}
\scriptsize
\setlength{\tabcolsep}{3.1pt}
\renewcommand{\arraystretch}{1.08}
\begin{tabular}{p{3.05cm}p{2.45cm}p{2.55cm}p{2.15cm}p{2.65cm}p{3.25cm}}
\hline
\textbf{System} & \textbf{Spatial filter} & \textbf{Semantic model} & \textbf{Output} & \textbf{Update model} & \textbf{Recall/completeness evidence} \\
\hline
Lee et al.~\cite{lee2015mainmemoryspatialkeyword} & 2-D point radius & Boolean keywords & All & Not central & Perfect result recall \\
Tampakis et al.~\cite{tampakis2021spatialkeywordrange} & 2-D radius & Jaccard threshold & All & Static & Exact filter--refine \\
Oh et al.~\cite{oh2018word2vecrange} & 2-D range & Word2Vec & Range results & Moving objects & No query certificate established \\
WISK~\cite{sheng2023wisk} & 2-D region & Boolean keywords & All & Buffered inserts; local retraining & Exact keyword processing \\
Dong et al.~\cite{dong2021continuoustopk} & Spatial proximity & Keyword relevance & Top-$k$ & Dynamic objects & Maintained top-$k$ semantics \\
LIST~\cite{yin2025list} & Learned spatial relevance & Dense embeddings & Top-$k$ & Insert; delete; index-only retraining & Empirical top-$k$ effectiveness \\
Mesh~\cite{song2025mesh} & 2-D spatial range & Dense vectors & Approximate top-$k$ & Static & Empirical ANN recall \\
Vector-radius retrieval~\cite{manohar2025rangeretrieval} & None & Vector-distance radius & Approximate all & Static & Empirical reporting recall \\
SeRF~\cite{zuo2024serf} & Scalar interval & Dense vectors & Approximate top-$k$ & Static & Empirical ANN recall \\
iRangeGraph~\cite{xu2024irangegraph} & Scalar interval & Dense vectors & Approximate top-$k$ & Static & Empirical ANN recall \\
Dynamic RFANNS~\cite{peng2025dynamicrfann} & Scalar interval & Dense vectors & Approximate top-$k$ & Dynamic & Empirical ANN recall \\
DIGRA~\cite{jiang2025digra} & Scalar interval & Dense vectors & Approximate top-$k$ & Dynamic graph & Empirical ANN recall \\
RangePQ~\cite{zhang2025rangepq} & Scalar interval & Dense vectors & Approximate top-$k$ & Dynamic quantization index & Empirical ANN recall \\
WoW~\cite{wang2025wow} & Scalar window & Dense vectors & Approximate top-$k$ & Incremental insert-from-empty & Empirical ANN recall \\
\textbf{FRESH-GEORANGE} & \textbf{2-D geodesic radius} & \textbf{Dense threshold} & \textbf{All / certified} & \textbf{Versioned insert, delete, replace} & \textbf{Exact mode or deterministic per-query lower bound} \\
\hline
\end{tabular}
\end{table*}

\textbf{Filtered and range-reporting vector search.}
Filtered-DiskANN demonstrates graph search with categorical filters~\cite{gollapudi2023filtereddiskann}; SeRF, iRangeGraph, UNIFY, and window-filter methods exploit ordered filter structure~\cite{zuo2024serf,xu2024irangegraph,liang2024unify,engels2024windowfilters}. More recent work makes the filtered index mutable: Dynamic RFANNS, DIGRA, and RangePQ address changing data~\cite{peng2025dynamicrfann,jiang2025digra,zhang2025rangepq}, while WoW is a fully incremental insert-from-empty design but leaves frequent deletion and in-place update as open concerns~\cite{wang2025wow}. These methods inform graph routing, segment organization, and update engineering. Their conventional query asks for the nearest $k$ vectors satisfying a one-dimensional range, however; adapting longitude or a space-filling key to that interface does not faithfully represent a circular geographic predicate. Manohar et al. instead study approximate reporting of every vector within a vector-distance radius~\cite{manohar2025rangeretrieval}, which is the direct threshold-output precedent, but lacks a geographic constraint, snapshot visibility, and deterministic accounting of missed objects.

\textbf{Freshness, drift, and guarantees.}
HNSW supplies the graph-search foundation used by many ANN systems~\cite{malkov2020hnsw}, but benchmark recall remains an empirical aggregate. SPFresh performs incremental in-place vector-index updates~\cite{xu2023spfresh}; VStream addresses distributed streaming search~\cite{gong2025vstream}; and Quake adapts partitioning and predicts parameters for requested recall levels under changing workloads~\cite{mohoney2025quake}. These systems motivate delta maintenance and drift-triggered consolidation, not the certificate itself. ConANN provides an important formal counterpoint by controlling ANN risk with distribution-free conformal methods~\cite{horchidan2025conann}. Its statistical guarantee and assumptions differ from a deterministic statement about one radius-and-threshold result set. In FRESH-GEORANGE, admissible block bounds and a conservative count of every unresolved visible record yield the certificate; learned routing and ANN traversal may reduce work but are never trusted as proof. Table~\ref{tab:closest-work} summarizes why no individual ingredient is unprecedented and why the proposed contribution must be evaluated as a conjunction.

\section{Method}
\label{sec:method}

\subsection{Problem Formulation}

Unlike top-$k$ search, our threshold query must return every live object that
passes its spatial, semantic, and freshness predicates.  Define
\begin{equation}
 \begin{aligned}
 x_i&=(i,\nu_i,p_i,v_i,a_i,o_i),\\
 Q&=(\ell,r,q,\tau,t_q,A_{\max},\Delta_{\max},\rho_{\rm req}).
 \end{aligned}
 \label{eq:record-query}
\end{equation}
where $p_i$ is a location, $v_i\in\mathbb{R}^d$ a unit embedding, $a_i$ an
observation day, and $o_i$ an upsert or delete (whose non-key fields may be
absent).  Sequence $\nu_i$ totally orders an identifier's events.  Query $Q$
specifies center, radius $r\geq0$, unit vector, inclusive cosine threshold
$\tau\in[-1,1]$, time, maximum record age and watermark lag, and requested
certificate $\rho_{\rm req}\in[0,1]$ (Table~\ref{tab:notation}).

\begin{table*}[t]
\caption{Notation.  Time is represented by integer replay day in the pilot.}
\label{tab:notation}
\centering
\footnotesize
\begin{tabular}{@{}cl@{\hspace{1.7em}}cl@{}}
\toprule
Symbol & Meaning & Symbol & Meaning \\
\midrule
$i$ & stable object identifier & $\nu_i$ & totally ordered event sequence \\
$p_i,\ell$ & object and query locations & $v_i,q$ & unit object and query vectors \\
$d$ & embedding dimension & $r$ & inclusive range radius \\
$\tau$ & inclusive cosine threshold & $t_q$ & query day \\
$a_i$ & object observation day & $A_{\max}$ & maximum object age \\
$W$ & committed source watermark & $\Delta_{\max}$ & maximum watermark lag \\
$B,D_W$ & immutable base and current delta & $\mathcal{L}_W$ & live view at $W$ \\
$C,b$ & spatial cell and semantic block & $L_C^p$ & spatial lower bound \\
$U_b^s,U_b^a$ & semantic and age upper bounds & $\mathcal{U}$ & unexamined, possible blocks \\
$\widehat A$ & verified returned identifiers & $M_{\rm rem}$ & residual physical-slot count \\
$\rho_{\rm req}$ & requested certified recall & $D_t$ & drift statistic \\
\bottomrule
\end{tabular}
\end{table*}

Exact verification uses great-circle distance, cosine similarity, and age;
exponential freshness is only a monotone reparameterization:
\begin{equation}
 \begin{aligned}
 d_i&=d_{\rm geo}(p_i,\ell),& s_i&=q^\top v_i,\\
 g_i&=(t_q-a_i)_+,& \phi_i(t_q)&=\exp(-\lambda g_i).
 \end{aligned}
 \label{eq:primitive}
\end{equation}
The implementation uses $g_i\leq A_{\max}$, equivalent for $\lambda>0$ to
$\phi_i\geq\exp(-\lambda A_{\max})$, but never multiplies similarity by decay.
Record age and source currency obey distinct contracts:
\begin{equation}
 0\leq t_q-W\leq\Delta_{\max},\qquad g_i\leq A_{\max}.
 \label{eq:freshness-contract}
\end{equation}
The first concerns completeness of the captured source prefix; the second
filters a stored observation day.  Neither proves recent physical inspection.
Pilot observation/event days are deterministic experimental attributes, not
historical OSM edits, so freshness claims apply only to the replay.

At watermark $W$, the range-complete answer is
\begin{equation}
 A_W^*(Q)=\{i\in\mathcal L_W:\ d_i\leq r,\ s_i\geq\tau,
 \ t_q-a_i\leq A_{\max}\}.
 \label{eq:answer}
\end{equation}
All predicates are inclusive; subsequent ranking cannot alter membership.

\subsection{Index and Current-View Semantics}

FRESH-GEORANGE separates an immutable base from a mutable replacement delta.
For the general event model, latest-write-wins semantics are
\begin{equation}
 \begin{aligned}
 e_i^W&=\underset{e\in B\cup D_{\leq W}:e.i=i}{\arg\max}\ e.\nu,\\
 \mathcal L_W&=\{e_i^W:e_i^W.o=\mathsf{upsert}\}.
 \end{aligned}
 \label{eq:lww}
\end{equation}
The pilot is narrower: \texttt{apply\_events} replays the complete prefix to a
private one-value-per-ID dictionary, whose final sequence-ordered value is an
upsert or tombstone.  It atomically publishes that dictionary and $W$ under a
re-entrant lock; queries hold the lock throughout.  Dictionary membership
hides an old base copy.  Building privately leaves readers on the preceding
complete state, while publishing $W$ before the dictionary would invalidate
the proof.  Thus calls are serializable, but there is no retained MVCC,
lock-free reading, or durable recovery: $t_q<W$ and
$t_q-W>\Delta_{\max}$ are rejected.  Past-prefix republication is an evaluation
operation, not a historical query.  Inputs must have complete, deterministic
sequences.  The pilot generator produces a unique, gap-free sequence; malformed
or conflicting external sequences are outside the implemented contract.

\begin{figure*}[t]
\centering
\resizebox{0.88\textwidth}{!}{\begin{tikzpicture}[
  x=1cm,y=1cm,>=stealth,
  box/.style={draw,rounded corners=2pt,fill=blue!5,align=center,
              minimum height=0.78cm,text width=2.8cm,inner sep=4pt},
  state/.style={box,fill=orange!11},
  work/.style={box,fill=green!8},
  result/.style={box,fill=blue!12},
  monitor/.style={box,fill=gray!10,dashed},
  flow/.style={->,line width=0.65pt},
  feedback/.style={->,dashed,line width=0.65pt},
  every node/.style={font=\scriptsize}
]
  \node[box] (base) at (-4.5,0) {Pinned day-30\\base snapshot};
  \node[state] (events) at (0,0) {Complete replay prefix\\of events $E_{\leq W}$};
  \node[box] (query) at (4.5,0) {Range query\\$(\ell,r,q,\tau)$; $(t_q,A_{\max},\Delta_{\max})$};

  \node[box] (blocks) at (-5.5,-1.7) {$10^\circ$ cells and\\semantic micro-blocks};
  \node[box] (graph) at (-2.0,-1.7) {Exact-$k$NN graph\\approximate proposer};
  \node[state] (builder) at (1.0,-1.7) {Private full-prefix\\delta reconstruction};
  \node[work] (contract) at (4.6,-1.7) {Watermark-lag check\\record-age check};

  \node[box,text width=3.1cm] (baseview) at (-3.8,-3.4)
    {Immutable base view\\arrays + summaries\\exact-$k$NN proposer};
  \node[state] (publish) at (0,-3.4) {Atomic locked publication\\of $(D_W,W)$};
  \node[work,text width=4.1cm] (capture) at (0,-4.9)
    {One locked current query view: base index, $(D_W,W)$, and validated $Q$};

  \node[work] (prune) at (-4.2,-6.5) {Safe base pruning\\spatial + semantic + age};
  \node[work] (delta) at (0,-6.5) {Exact scan of every\\delta value};
  \node[work] (verify) at (4.2,-6.5) {Ordered exact base\\verification (graph first)};

  \node[result,text width=3.5cm] (merge) at (0,-8.1)
    {Deduplicated verified IDs and physical residual count};
  \node[monitor] (drift) at (-4.5,-9.8) {Scripted JSD monitor\\day 50: rebuild if\\JSD $>0.05$};
  \node[result] (exact) at (0,-9.8) {Exact set\\$M_{\rm rem}=0$, certificate $1$};
  \node[result] (cert) at (4.5,-9.8) {Certified set\\$\widehat A$ and lower bound $\underline R$};

  \draw[flow] (base) -- (blocks);
  \draw[flow] (base) -- (graph);
  \draw[flow] (blocks) -- (baseview);
  \draw[flow] (graph) -- (baseview);
  \draw[flow] (events) -- (builder);
  \draw[flow] (builder) -- (publish);
  \draw[flow] (query) -- (contract);
  \draw[flow] (baseview) -- (capture);
  \draw[flow] (publish) -- (capture);
  \draw[flow] (contract) -- (capture);
  \draw[flow] (capture) -- (prune);
  \draw[flow] (capture) -- (delta);
  \draw[flow] (capture) -- (verify);
  \draw[flow] (prune) -- (merge);
  \draw[flow] (delta) -- (merge);
  \draw[flow] (verify) -- (merge);
  \draw[feedback] (merge) -- (drift);
  \draw[flow] (merge) -- (exact);
  \draw[flow] (merge) -- (cert);
  \draw[feedback] (drift.west) -- (-6.6,-9.8) -- (-6.6,-3.4) -- (baseview.west);
\end{tikzpicture}}
\caption{FRESH-GEORANGE pilot architecture.  Solid arrows are implemented
data or query paths.  The dashed path is the scripted drift experiment: it
performs a full rebuild at a fixed checkpoint, not continuous retraining,
partial repartitioning, or the benefit-aware policy in Eq.~\eqref{eq:rebuild}.}
\label{fig:architecture}
\end{figure*}
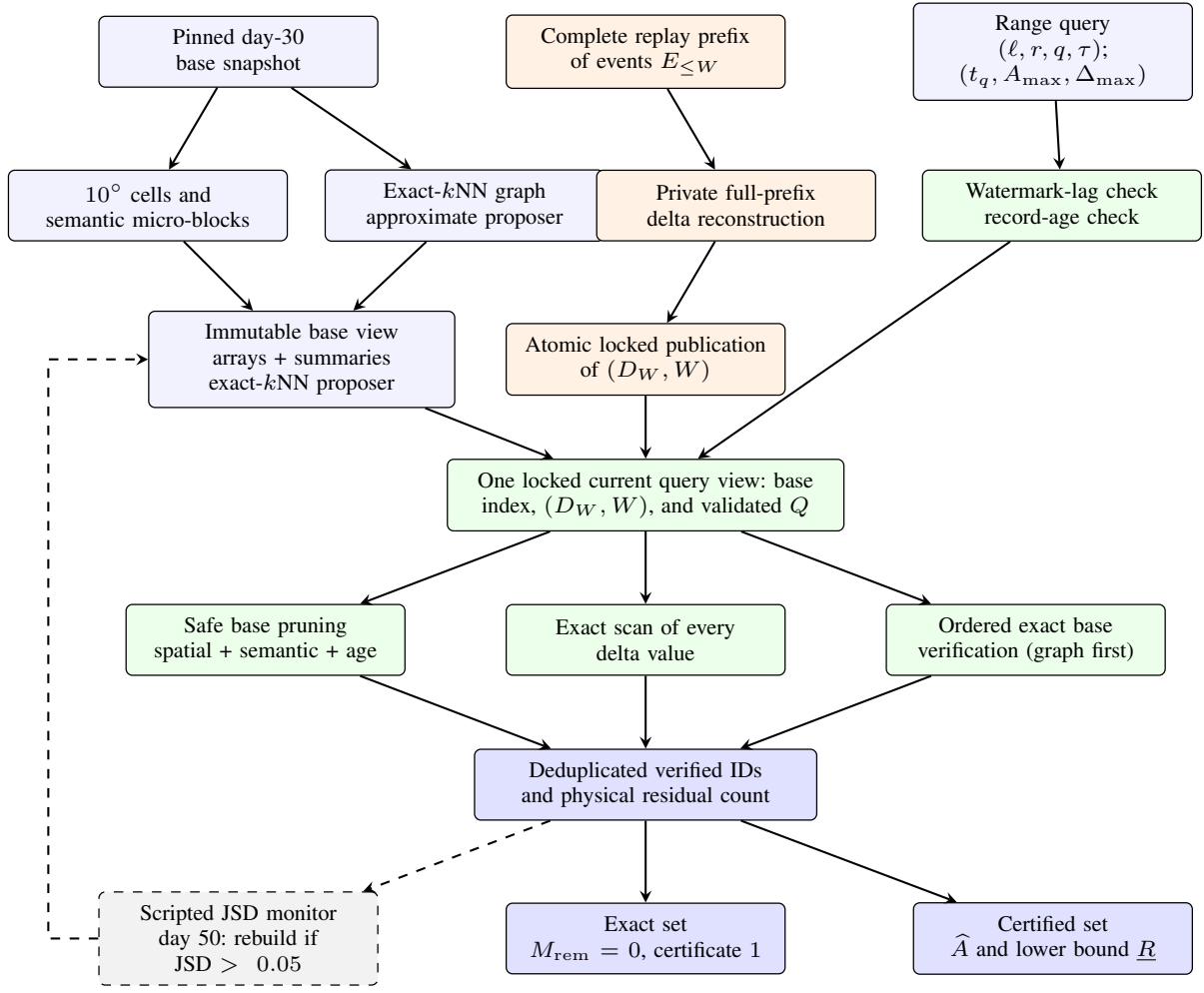

Figure~\ref{fig:architecture} separates the current-view path from scripted
drift feedback: the graph orders work; safe summaries and full delta scanning
support correctness.

The day-30 base alone fits a deterministic 48-D word/character TF--IDF--SVD
encoder.  Normalized vectors are grouped by fixed $10^\circ$ cells and
approximately 64-record semantic blocks; dense record arrays remain the
verification authority.  A brute-built exact-$k$NN graph is traversed
approximately only to propose base records---it is not HNSW or evidence of
absence.  Every current delta value is scanned, so unseen insertions remain
discoverable.  A replacement or tombstone hides its physical base slot; this
auditable design trades delta latency for safety.

For cell $C$, let $c_C^p$ be its center and let $\rho_C^p$ cover the complete
cell region under the declared great-circle metric.  The implementation uses
a meridian-then-parallel path with the parallel length evaluated at the
cell latitude closest to the equator.  The triangle inequality yields
\begin{equation}
 \begin{aligned}
 L_C^p(\ell)&=\max\{0,d_{\rm geo}(\ell,c_C^p)-\rho_C^p\},\\
 U_C^p(\ell)&=d_{\rm geo}(\ell,c_C^p)+\rho_C^p.
 \end{aligned}
 \label{eq:spatial-bound}
\end{equation}
Hence $L_C^p>r$ safely eliminates a cell, including near poles or the date line.

For each base block $b$, store the arithmetic center $c_b$, radius
$\rho_b=\max_{i\in b}\lVert v_i-c_b\rVert_2$, normalized center
$\mu_b=c_b/\lVert c_b\rVert_2$ when the norm is nonzero (one member vector is
used otherwise), and angular radius
$\theta_b=\max_{i\in b}\arccos(\mu_b^\top v_i)$.  Cauchy--Schwarz and the
spherical triangle inequality give two independent safe upper bounds:
\begin{equation}
 \begin{aligned}
 U_b^{\rm ball}(q)&=\min\{1,q^\top c_b+\rho_b\},\\
 \eta_b(q)&=\max\{0,\arccos(\operatorname{clip}
   (q^\top\mu_b,-1,1))-\theta_b\},\\
 U_b^{\rm ang}(q)&=\cos(\eta_b(q)),\\
 U_b^s(q)&=\min\{U_b^{\rm ball}(q),U_b^{\rm ang}(q)\}.
 \end{aligned}
 \label{eq:semantic-bound}
\end{equation}
Clip $\arccos$ arguments to $[-1,1]$.  Both components upper-bound every block
dot product, so their minimum is safe.  For
$a_b^{\max}=\max_{i\in b}a_i$, age requires
\begin{equation}
 U_b^a=a_b^{\max},\qquad U_b^a\geq t_q-A_{\max}.
 \label{eq:age-bound}
\end{equation}
A fixed outward margin protects bounds; verification uses float64.  Pruning is
only a disjunction of necessary-condition failures:
\begin{equation}
 \begin{aligned}
 \operatorname{prune}(b,Q)\iff{}&L_{C(b)}^p(\ell)>r\ \lor\ U_b^s(q)<\tau\\
 &{}\lor\ U_b^a<t_q-A_{\max}.
 \end{aligned}
 \label{eq:prune}
\end{equation}
No learned score prunes.  Immutable base blocks and a full delta scan avoid
dynamic radius expansion in this pilot.

For any $i\in b$,
$q^\top v_i=q^\top c_b+q^\top(v_i-c_b)\leq q^\top c_b+\rho_b$ because
$\lVert q\rVert_2=1$.  Spherical triangle inequality gives angle at least
$\angle(q,\mu_b)-\theta_b$; cosine decreases on $[0,\pi]$.  Metric triangle
inequality proves~\eqref{eq:spatial-bound}.  Therefore loose summaries admit
false positives but cannot prune true members, assuming normalized nonzero
vectors and the outward margin.  Sampled bound audits check implementation
maxima but do not replace this proof.

\subsection{Exact and Certified Execution}

Algorithm~\ref{alg:exact} describes exact mode as implemented.  Graph traversal
only changes the order of some checks; completion scans every unexamined slot
in every non-pruned base block.  Each candidate is checked for delta shadowing,
hard age, exact dot product, and exact great-circle distance.

\begin{algorithm}[t]
\caption{\textsc{Exact-Range}$(Q)$}
\label{alg:exact}
\small
\begin{algorithmic}[1]
\State acquire the index lock and capture $(B,D_W,W)$
\Require $t_q\geq W$ and $t_q-W\leq\Delta_{\max}$
\State encode and normalize $q$; compute non-pruned base blocks $\mathcal S$
\State $\widehat A\gets\varnothing$; $E\gets\varnothing$
\For{each value in $D_W$}
  \If{it is an upsert and \textsc{Verify} passes}
    \State add its identifier to $\widehat A$
  \EndIf
\EndFor
\For{each graph-proposed base slot in $\mathcal S$}
  \State mark the slot in $E$ and exactly verify it, including visibility
\EndFor
\For{each base slot in $\mathcal S\setminus E$}
  \State mark and exactly verify the slot; add every passing identifier
\EndFor
\State \Return sorted $\widehat A$, certificate $1$, and watermark $W$
\end{algorithmic}
\end{algorithm}

Certified mode scans all deltas, takes graph proposals, then visits surviving
blocks by decreasing semantic upper bound.  For examined base slots $E$ and
non-pruned blocks with unexamined slots $\mathcal U$, define
\begin{equation}
 M_{\rm rem}=\sum_{b\in\mathcal U}\left|\{j\in b:j\notin E\}\right|.
 \label{eq:residual}
\end{equation}
This physical count includes shadowed and predicate-failing records; such
overcounting only weakens the bound.  For exactly verified IDs $\widehat A$,
\begin{equation}
 \begin{aligned}
 \underline R&=
 \begin{cases}
 1,&|\widehat A|=0\ \wedge\ M_{\rm rem}=0,\\
 |\widehat A|/(|\widehat A|+M_{\rm rem}),&\text{otherwise},
 \end{cases}\\
 &\text{stop if }M_{\rm rem}=0\ \lor\ \underline R\geq\rho_{\rm req}.
 \end{aligned}
 \label{eq:certificate}
\end{equation}

\begin{algorithm}[t]
\caption{\textsc{Certified-Range}$(Q,\rho_{\rm req},b,e_f)$}
\label{alg:certified}
\small
\begin{algorithmic}[1]
\State capture and validate the locked current view
\State prune base blocks safely; exactly scan every current delta value
\State request at most $e_f$ graph proposals; exactly verify allowed slots
\State compute $M_{\rm rem}$ and $\underline R$
\While{$\underline R<\rho_{\rm req}$ and fewer than $b+e_f$ base slots are examined}
  \State choose the remaining block with largest $U_b^s(q)$
  \State scan its unexamined slots until the block or cap ends
  \State update $\widehat A$, $M_{\rm rem}$, and $\underline R$
\EndWhile
\State \Return sorted $\widehat A$, $\underline R$, $M_{\rm rem}$, and $W$
\end{algorithmic}
\end{algorithm}

The certificate assumes the locked snapshot, safe bounds, exact verification,
and normalized vectors, but no graph recall.  For missed answers
$\mathcal M=A_W^*(Q)\setminus\widehat A$, all deltas were checked and
Eq.~\eqref{eq:prune} excludes pruned blocks; each miss therefore maps to an
unexamined physical slot in~\eqref{eq:residual}:
\begin{equation}
 \begin{aligned}
 |\mathcal M|&\leq M_{\rm rem}\quad\Longrightarrow\\
 \operatorname{Recall}(\widehat A)
 &=\frac{|\widehat A|}{|\widehat A|+|\mathcal M|}\geq\underline R.
 \end{aligned}
 \label{eq:certificate-proof}
\end{equation}
Empty truth has recall one; the bound reaches one only when the residual is
zero, as exact mode ensures.  An unexamined shadowed slot conservatively remains
in $M_{\rm rem}$; examining it removes the slot without adding an answer.
Delta answers may enter the numerator because they were verified, and never
enter the residual because all deltas were scanned.  ID deduplication prevents
double counting.  Thus the certificate is a lower bound, not empirical-recall
estimation; tightness reflects how quickly pruning/scanning removes slots.

\subsection{Drift, Maintenance, and Cost Policy}

For a full deployment, fixed, smoothed histograms can track spatial occupancy,
semantic-cluster occupancy, query location, radius, and threshold.  With
reference distribution $P_{j,0}$, recent distribution $P_{j,t}$, and
$M_j=(P_{j,0}+P_{j,t})/2$, define
\begin{equation}
 \begin{aligned}
 D_t&=\sum_j\alpha_j\operatorname{JSD}(P_{j,t},P_{j,0}),
 &\sum_j\alpha_j&=1,\\
 \operatorname{JSD}(P_{j,t},P_{j,0})
 &=\tfrac12\operatorname{KL}(P_{j,t}\Vert M_j)\\
 &\quad+\tfrac12\operatorname{KL}(P_{j,0}\Vert M_j).
 \end{aligned}
 \label{eq:drift}
\end{equation}
A defensible production decision would require persistent drift and amortized
benefit:
\begin{equation}
 \begin{aligned}
 \operatorname{REBUILD}_t={}&
 \mathbf{1}\!\left[\min_{0\leq h<k}D_{t-h}>\gamma_D\right]\\
 &\cdot\mathbf{1}\!\left[H\widehat{\Delta L}_{p99}>C_{\rm rebuild}\right].
 \end{aligned}
 \label{eq:rebuild}
\end{equation}
Here $H$ is forecast query count and $\widehat{\Delta L}_{p99}$ conservative
per-query benefit in rebuild-cost units; compaction is separate.  The pilot
does \emph{not} implement Eq.~\eqref{eq:rebuild}: it uses a smoothed 24-bin
query-location histogram and one day-50 rebuild after a scripted shift if
JSD $>0.05$.  This checks mechanism, not online optimality or partial rebuilds.

More generally, a configuration $\pi$ could be selected with normalized costs
and hard guarantees:
\begin{equation}
 \begin{aligned}
 \min_{\pi}\quad &w_L\widehat L_{p99}+w_U\widehat C_{\rm update}
 +w_M\widehat M\\
 &{}+w_S\widehat{(t_q-W)}+w_B\widehat C_{\rm rebuild}\\
 \text{s.t.}\quad&\underline R_{\pi}\geq\rho_{\rm req},\quad
 t_q-W\leq\Delta_{\max},\quad M_{\pi}\leq M_{\max}.
 \end{aligned}
 \label{eq:objective}
\end{equation}
Hats denote division by preregistered reference scales.  Equation~\eqref{eq:objective}
uses $M_\pi$ for memory footprint (not $M_{\rm rem}$) and is a design
objective, not an optimizer implemented or tuned in this pilot.

\subsection{Complexity, Limits, and Edge Cases}

For $m$ surviving blocks, $n_{\rm scan}$ examined base slots, delta size
$|D_W|$, and graph degree $g$, query work is
$O(md+n_{\rm scan}d+|D_W|d)$ plus at most $e_f$ vertices' graph work; worst-case
exact search is linear.  Storage is $O((N+|D_W|)d+Ng)$, including duplicated
graph vectors.  Brute-force neighbor construction can be $O(N^2d)$: no HNSW
or sublinear guarantee is claimed.  Full-prefix replay and embedding makes a
pilot publication $O(|E_{\leq W}|d)$, not a production append-only $O(d)$;
reported update throughput is implementation-specific.

Boundary equality is resolved only by the exact inclusive predicates.  A zero
radius returns co-located matches; $\tau=-1$ disables semantic exclusion, while
$\tau=1$ is checked in float64.  Great-circle distance and wrapped longitude
handle date-line and polar queries.  Multiple events for an identifier resolve
by final sequence order; a tombstone hides all base state.  An exact empty
answer has certificate one, whereas an early empty answer with residual slots
has certificate zero.  Unknown-region delta objects remain safe because the
delta is not spatially pruned.  Missing trustworthy observation time should
disable the object-age predicate rather than substitute last-edit time.  The
method assumes finite embeddings in one fixed model generation; the current
prototype normalizes nonzero outputs, treats a zero vector as having zero
cosine with every query, and does not implement
dual-generation migration.  These restrictions delimit the claims evaluated
in Section~\ref{sec:experiments}.

\section{Experimental Evaluation}
\label{sec:experiments}

This section reports a functional CPU pilot, not a publication-scale comparison.
The narrow purpose is to check the query semantics, exact mode, certificate
accounting, and update-overlay plumbing on an executable workload.  It does not
substitute for the proposed OSM-diff evaluation, and the IVF, graph, lexical,
and LIST-router implementations are local reference proxies rather than the
authors' systems.  Results are therefore interpreted within this pilot only.

\subsection{Experimental Setup}

\paragraph{Data and replay.}
We pinned a 2,500-row airport sample from the OpenFlights-format airport
file~\cite{openflights2026data}; every retained row declares OurAirports, whose
official page releases the data to the Public Domain~\cite{ourairports2026data},
as its source (SHA-256
\texttt{0ad06d\ldots 44cd}).  Its names, cities, countries, facility fields,
and coordinates are source attributes.  We used the first 2,000 rows as the
day-30 base and held out 500 real
rows for insertions on days 31--60.  A seeded generator added 160 metadata
updates and 80 deletions, yielding 740 replay events and 2,420 live objects on
day 60.  Event types, event days, and \texttt{observed\_day} are simulated, so
this experiment tests a simulated age predicate and source-watermark behavior;
it provides no evidence about historical airport changes or real
record-freshness distributions.  Table~\ref{tab:data} records the snapshots.

\begin{table}[t]
\centering
\caption{Pilot data snapshots. Object attributes are real; replay operations and times are simulated.}
\label{tab:data}
\small
\begin{tabular}{lrrr}
\toprule
Snapshot & Objects & Countries & Events \\
\midrule
Pinned raw sample & 2500 & 155 & 0 \\
Base & 2000 & 132 & 0 \\
Live after replay & 2420 & 154 & 740 \\
\bottomrule
\end{tabular}
\end{table}

Table~\ref{tab:configuration} records the resolved, executable settings rather
than an intended configuration. All later plot data and prose macros are
exported from the matching run.

\begin{table}[t]
\centering
\caption{Resolved pilot configuration. Values are exported from the YAML files and execution manifest.}
\label{tab:configuration}
\scriptsize
\begin{tabularx}{\columnwidth}{@{}lX@{}}
\toprule
Setting & Resolved value \\
\midrule
Raw/base/live objects & 2,500 / 2,000 / 2,420 \\
Simulated replay events & 740 \\
Primary seeds & 101, 211, 307, 401, 503 \\
Queries per seed / total & 36 / 180 \\
Radii (km) & 150, 500, 1500, 4000 \\
Cosine thresholds & 0.15, 0.25, 0.35, 0.45 \\
Maximum simulated ages (days) & 14, 30, 60, 60 \\
Encoder / dimensions & TF--IDF+LSA / 48 \\
Cell / block size & 10.0 deg / 64 records \\
Graph degree / proposal budget & 12 / 64 \\
Certificate target & 0.95 \\
CPU / logical CPUs & AMD EPYC 9V74 80-Core Processor / 9 \\
Python / scikit-learn & 3.12.14 / 1.8.0 \\
\bottomrule
\end{tabularx}
\end{table}

\paragraph{Representation and index settings.}
A deterministic word-and-character TF--IDF pipeline followed by 48-dimensional
truncated SVD produced unit-normalized vectors.  It was fit on the day-30 base
only; insertion and update text was unseen during fitting.  FRESH-GEORANGE
used $10^\circ$ spatial cells, semantic blocks of at most 64 records, a
degree-12 exact-$k$NN reference graph used only for proposal order, and search
budget 64.  Certified queries requested a 0.95 lower bound; exact queries
continued verification to completion.  The query driver applied the complete
event history through a watermark before querying.  This replacement-batch API
assumes one writer and a complete history; it is not a retained, concurrent
micro-batch implementation.

\paragraph{Queries, comparators, and ground truth.}
For each of five seeds (101, 211, 307, 401, and 503), we sampled 36 day-60
queries from the live corpus, for 180 distinct queries.  A query center is a
small perturbation of a sampled airport and its phrase is formed from that
row's derived facility kind, city, and country.  Radii rotate over 150, 500,
1,500, and 4,000~km; cosine thresholds over 0.15, 0.25, 0.35, and 0.45; and
maximum simulated ages over 14, 30, 60, and 60 days.  These broad radii and
source-derived phrases make this a predicate smoke test, not an independently
judged emergency-resource workload.

Brute-force, spatial-first, and semantic-first exact scans establish set-valued
ground truth.  We additionally tested a lexical spatial grid, IVF post-filter,
an exact-$k$NN graph post-filter, a LIST-style cluster router, and a spatial
index rebuilt every 15 days, plus FGR-Cert95 and FGR-Exact.  The local IVF,
graph, lexical, and LIST-style implementations neither reproduce HNSW, LIST,
nor an IR-tree nor justify external superiority claims.  Ten methods over 180
queries produced \PrimaryRuns{} method--query executions.

\paragraph{Measurement and statistics.}
For truth $T$ and output $A$, recall, precision, and Jaccard are
$|A\cap T|/|T|$, $|A\cap T|/|A|$, and $|A\cap T|/|A\cup T|$ (empty--empty is
one). Higher set scores and throughput are better; lower latency, staleness,
footprint, and rebuild cost are better.
Method order was randomized within each measured query, but every query--method
pair was timed only once.  Latency is therefore exploratory single-query
wall-clock time.  The
driver issued one query at a time on a CPU environment exposing
\PilotLogicalCPUs{} logical CPUs: \PilotCPU{}, Linux \PilotKernel{}, Python
\PilotPython{}, NumPy \PilotNumpy{}, pandas \PilotPandas{}, scikit-learn
\PilotSklearn{}, and SciPy \PilotScipy{}.  Peak process RSS was
\PilotPeakRSSKiB{}~KiB.
Footprint is protocol-5 serialized object size, not resident memory.
``Examined'' is meaningful within a method but is not comparable across
methods because the implementations count different operations.

For inferential summaries, the five independently constructed seeds---not the
180 dependent query rows---are the paired units.  We applied two-sided Wilcoxon
signed-rank tests and Holm adjustment within each metric; intervals in
Table~\ref{tab:stats} bootstrap the five paired seed differences.  The pooled
query-row recall intervals in the main summary are not used for inference.
With $n=5$, the tests are coarse and underpowered and are included
as an uncertainty check rather than evidence of equivalence or superiority.

\subsection{Primary Result: Correctness at a Latency Cost}

Table~\ref{tab:main_results} gives the principal result.  FGR-Exact attained
\FGRExactRecall{} recall, precision, and Jaccard on all
\FGRExactPerfectQueries{} primary queries.  FGR-Cert95 attained
\FGRCertRecall{} mean empirical recall and \FGRCertPrecision{} precision while reporting a
\FGRCertBound{} mean lower bound.  The mean empirical-minus-certified gap was
only \FGRCertGap; the certificate-violation count was
\FGRCertViolationCount{} of \QueriesPerMethod{}, the minimum certificate was
\FGRCertMinBound{}, and the below-target count was
\FGRCertBelowTargetCount{}.  Across five seeds, a separate audit made
\BoundAuditComparisons{} sampled semantic-bound comparisons; the smallest
slack was positive (\BoundAuditMinSlack{}).  Ten unit
tests also covered exact/certificate behavior, normalization, terminal cells,
non-divisor grids, poles, the dateline, singleton graphs, and watermark rules.
These checks support the certificate on the declared path---normalized vectors,
a committed snapshot, complete-history replay, and one ordered writer---but are
not a general production proof.  Repeated versions, moves, late or gapped
events, concurrent writers, malformed embeddings, and semantic outlier streams
remain outside the property suite.

\begin{table*}[t]
\centering
\caption{Executed day-60 pilot results across five seeds and 180 queries per method. Recall and precision are percentages. Footprint is standalone pickle size, not resident memory.}
\label{tab:main_results}
\small
\setlength{\tabcolsep}{4.2pt}
\begin{tabular}{lrrrrrr}
\toprule
Method & Recall $\uparrow$ & Precision $\uparrow$ & Median ms $\downarrow$ & p95 ms $\downarrow$ & Examined & Serialized MB \\
\midrule
Brute-force exact & 100.00 & 100.00 & 1.39 & 2.03 & 2420.0 & 5.07 \\
Spatial-first exact & 100.00 & 100.00 & 1.24 & 1.99 & 265.4 & 5.18 \\
Semantic-first exact & 100.00 & 100.00 & 1.23 & 1.86 & 448.5 & 5.07 \\
Lexical spatial & 99.96 & 100.00 & 1.44 & 2.76 & 265.4 & 5.40 \\
IVF-post & 85.45 & 97.78 & 1.29 & 1.84 & 592.9 & 5.12 \\
kNN-graph post-filter & 32.56 & 55.56 & 2.31 & 3.19 & 96.0 & 5.18 \\
LIST-router proxy & 85.59 & 98.33 & 1.27 & 1.79 & 504.8 & 5.11 \\
Periodic rebuild (15 d) & 68.01 & 78.20 & 1.28 & 2.14 & 257.9 & 5.04 \\
FGR-Cert95 (this work) & 99.91 & 100.00 & 7.24 & 12.39 & 1047.4 & 5.39 \\
FGR-Exact (this work) & 100.00 & 100.00 & 7.27 & 12.53 & 1048.5 & 5.39 \\
\bottomrule
\end{tabular}
\end{table*}

The efficiency result is negative.  FGR-Cert95 required a
\FGRCertMedianMs{}-ms median and \FGRCertPNinetyFiveMs{}-ms p95, versus
\SpatialMedianMs{}~ms and \SpatialPNinetyFiveMs{}~ms for spatial-first exact---a
\LatencyRatio$\times$ median slowdown (\PNinetyFiveLatencyRatio$\times$ at p95).  It was also
essentially tied with FGR-Exact (\FGRExactMedianMs{}~ms), saving only
\FGRExaminedSaving{} reported examinations on average.  On this small corpus, certification and
overlay bookkeeping dominate, and early termination provides no meaningful
latency win.  Serialized footprint was \FGRSerializedMB{}~MB versus
\SpatialSerializedMB{}~MB
for spatial-first.  The correct conclusion is feasibility of the narrow
correctness contract, not a performance advantage.

\begin{figure*}[t]
  \centering
  \resizebox{\textwidth}{!}{\begin{tikzpicture}
\definecolor{fgblue}{RGB}{0,114,178}\definecolor{fgorange}{RGB}{213,94,0}\definecolor{fggreen}{RGB}{0,158,115}\definecolor{fgpurple}{RGB}{204,121,167}
\begin{groupplot}[group style={group size=2 by 1,horizontal sep=1.1cm},width=0.47\textwidth,height=0.25\textwidth,grid=major,tick label style={font=\scriptsize},label style={font=\footnotesize},legend style={font=\scriptsize,draw=none}]
\nextgroupplot[xlabel={Median latency (ms)},ylabel={Recall (\%)},title={(a) Recall--latency}]
\addplot+[fgblue,mark=*] table[x=latency_ms,y expr=100*\thisrow{recall},col sep=comma]{figures/data/frontier_fgr.csv};\addlegendentry{FGR}
\addplot+[fgorange,dashed,mark=square*] table[x=latency_ms,y expr=100*\thisrow{recall},col sep=comma]{figures/data/frontier_ivf.csv};\addlegendentry{IVF proxy}
\addplot+[fggreen,dotted,mark=triangle*] table[x=latency_ms,y expr=100*\thisrow{recall},col sep=comma]{figures/data/frontier_graph.csv};\addlegendentry{Graph proxy}
\addplot+[fgpurple,dashdotted,mark=diamond*] table[x=latency_ms,y expr=100*\thisrow{recall},col sep=comma]{figures/data/frontier_list.csv};\addlegendentry{LIST proxy}
\nextgroupplot[xlabel={Requested certificate},ylabel={Percent},title={(b) Certificate tightness},legend pos=south east]
\addplot+[fgblue,mark=*] table[x=target,y expr=100*\thisrow{recall},col sep=comma]{figures/data/frontier_fgr.csv};\addlegendentry{Empirical}
\addplot+[fgorange,dashed,mark=square*] table[x=target,y expr=100*\thisrow{certificate},col sep=comma]{figures/data/frontier_fgr.csv};\addlegendentry{Certified}
\end{groupplot}\end{tikzpicture}}
  \caption{Recall--latency and certificate frontier.  This is a
  single-query CPU pilot over 2,500 airport rows with a simulated 740-event
  replay; IVF and graph lines are local proxies, not official implementations.}
  \label{fig:recall-latency}
\end{figure*}
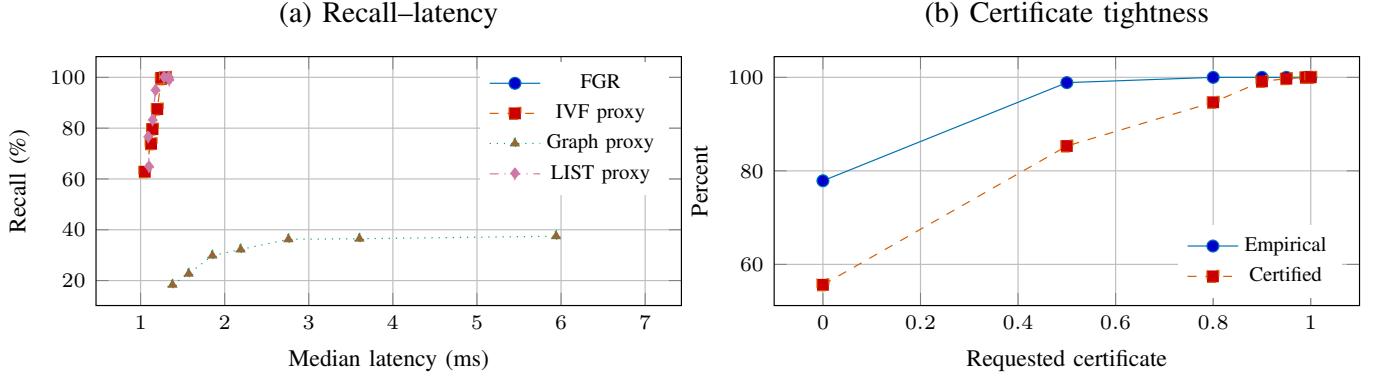

Figure~\ref{fig:recall-latency} reaches the same conclusion across budgets.
Increasing the FGR request from 0 to 0.95 raised empirical recall from
\FrontierFGRZeroRecall{} to \FrontierFGRNinetyFiveRecall{} and its reported
bound from \FrontierFGRZeroCert{} to \FrontierFGRNinetyFiveCert{}; median
latency remained between \FrontierFGRMinLatencyMs{} and
\FrontierFGRMaxLatencyMs{}~ms because the corpus is tiny and timing noise masks
a monotone work trend.  The IVF proxy reaches \FrontierIVFMaxRecall{} only when
all 32 clusters are probed, whereas the graph proxy reaches just
\FrontierGraphMaxRecall{} at its largest budget.
The plotted proxy curves measure these local implementations only and provide
no comparison to official ANN systems.

Table~\ref{tab:stats} reinforces the limits on comparative interpretation.
The estimated FGR-Cert95 recall differences were $\StatsDiffLIST{}$ versus the
LIST-router proxy, $\StatsDiffIVF{}$ versus IVF, and $\StatsDiffGraph{}$ versus the
graph proxy, but every Holm-adjusted $p$-value was at least
\StatsMinHolmP{}.  Its difference from FGR-Exact was $\StatsDiffExact{}$.  Five
seed pairs cannot sustain a strong performance claim.

\begin{table}[t]
\centering
\caption{Paired recall differences: FGR-Cert95 minus comparator. CIs are bootstrap 95\%; $p_H$ is Holm-adjusted Wilcoxon.}
\label{tab:stats}
\scriptsize
\begin{tabular}{lrrr}
\toprule
Comparator & Mean diff. & 95\% CI & $p_H$ \\
\midrule
LIST-router proxy & 0.143 & [0.105, 0.194] & 0.56 \\
FGR-Exact & -0.001 & [-0.001, -0.000] & 0.62 \\
Periodic rebuild (15 d) & 0.319 & [0.280, 0.352] & 0.56 \\
Spatial-first exact & -0.001 & [-0.001, -0.000] & 0.62 \\
IVF-post & 0.145 & [0.117, 0.174] & 0.56 \\
kNN-graph post-filter & 0.673 & [0.631, 0.709] & 0.56 \\
\bottomrule
\end{tabular}
\end{table}

\subsection{Sensitivity, Freshness, and Replay Cost}

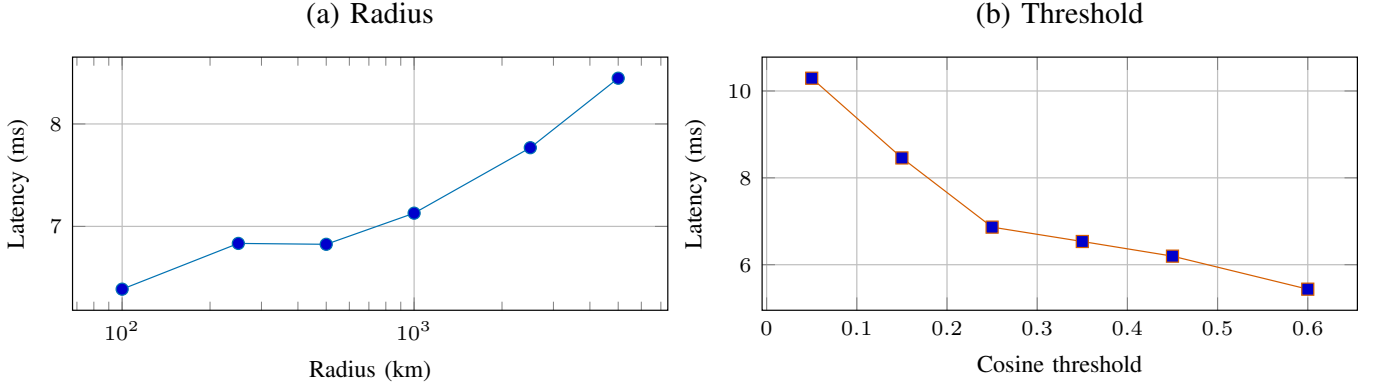
\begin{figure*}[t]
  \centering
  \resizebox{\textwidth}{!}{\begin{tikzpicture}\definecolor{fgblue}{RGB}{0,114,178}\definecolor{fgorange}{RGB}{213,94,0}\definecolor{fggreen}{RGB}{0,158,115}\definecolor{fgpurple}{RGB}{204,121,167}
\begin{groupplot}[group style={group size=2 by 1,horizontal sep=1.1cm},width=0.47\textwidth,height=0.25\textwidth,grid=major,tick label style={font=\scriptsize},label style={font=\footnotesize}]
\nextgroupplot[xmode=log,xlabel={Radius (km)},ylabel={Latency (ms)},title={(a) Radius}]
\addplot+[fgblue,mark=*] table[x=radius_km,y=latency_ms,col sep=comma]{figures/data/radius_sensitivity.csv};
\nextgroupplot[xlabel={Cosine threshold},ylabel={Latency (ms)},title={(b) Threshold}]
\addplot+[fgorange,mark=square*] table[x=threshold,y=latency_ms,col sep=comma]{figures/data/threshold_sensitivity.csv};
\end{groupplot}\end{tikzpicture}}
  \caption{Radius and threshold sensitivity in the single-query, 2,500-row
  airport pilot with simulated updates.  Points are aggregates over fixed
  20-query cohorts; they are not population-level estimates.}
  \label{fig:sensitivity}
\end{figure*}

Radius and threshold sweeps in Fig.~\ref{fig:sensitivity} reuse a fixed cohort
within each panel.  At radii through 1,000~km, both recall and certificate were
\RadiusOneKRecall{} and \RadiusOneKCert{}, respectively; at 5,000~km, latency
rose to \RadiusFiveKLatencyMs{}~ms, recall was \RadiusFiveKRecall{}, and the
bound was \RadiusFiveKCert{}.  Tightening the cosine threshold from 0.05 to
0.60 reduced mean answer count from \ThresholdLowAnswers{} to
\ThresholdHighAnswers{} and latency from \ThresholdLowLatencyMs{} to
\ThresholdHighLatencyMs{}~ms; recall rose from \ThresholdLowRecall{} to
\ThresholdHighRecall{}, while the bound rose from \ThresholdLowCert{} to
\ThresholdHighCert{}.  This is
consistent with fewer qualifying records, but only for this cohort and encoder.

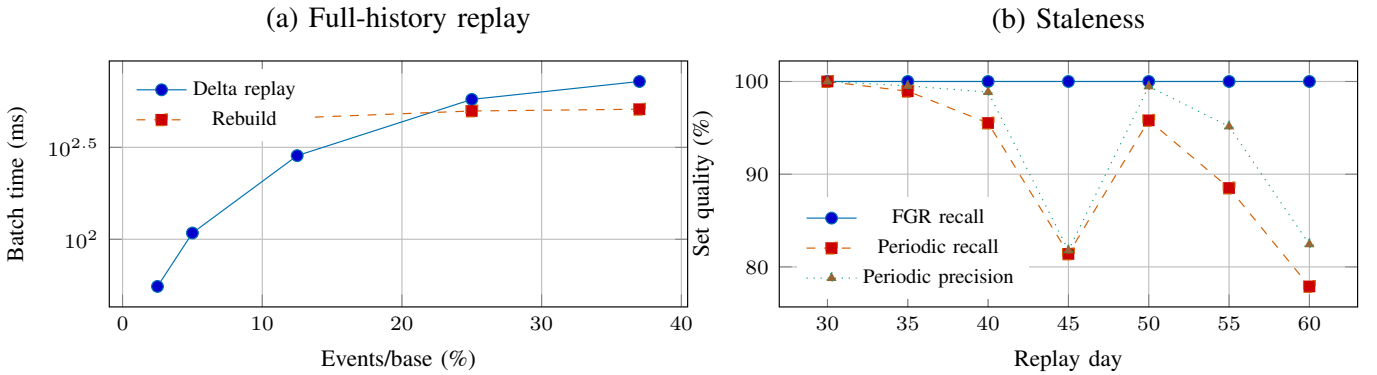
\begin{figure*}[t]
  \centering
  \resizebox{\textwidth}{!}{\begin{tikzpicture}\definecolor{fgblue}{RGB}{0,114,178}\definecolor{fgorange}{RGB}{213,94,0}\definecolor{fggreen}{RGB}{0,158,115}\definecolor{fgpurple}{RGB}{204,121,167}
\begin{groupplot}[group style={group size=2 by 1,horizontal sep=1.1cm},width=0.47\textwidth,height=0.25\textwidth,grid=major,tick label style={font=\scriptsize},label style={font=\footnotesize},legend style={font=\scriptsize,draw=none}]
\nextgroupplot[ymode=log,xlabel={Events/base (\%)},ylabel={Batch time (ms)},title={(a) Full-history replay},legend pos=north west]
\addplot+[fgblue,mark=*] table[x=update_percent,y=full_history_replay_ms,col sep=comma]{figures/data/update_scalability.csv};\addlegendentry{Delta replay}
\addplot+[fgorange,dashed,mark=square*] table[x=update_percent,y=full_rebuild_ms,col sep=comma]{figures/data/update_scalability.csv};\addlegendentry{Rebuild}
\nextgroupplot[xlabel={Replay day},ylabel={Set quality (\%)},title={(b) Staleness},legend pos=south west]
\addplot+[fgblue,mark=*] table[x=day,y expr=100*\thisrow{fgr_recall},col sep=comma]{figures/data/staleness_wide.csv};\addlegendentry{FGR recall}
\addplot+[fgorange,dashed,mark=square*] table[x=day,y expr=100*\thisrow{periodic_recall},col sep=comma]{figures/data/staleness_wide.csv};\addlegendentry{Periodic recall}
\addplot+[fggreen,dotted,mark=triangle*] table[x=day,y expr=100*\thisrow{periodic_precision},col sep=comma]{figures/data/staleness_wide.csv};\addlegendentry{Periodic precision}
\end{groupplot}\end{tikzpicture}}
  \caption{Full-history replay cost and 15-day rebuild staleness.  Both panels
  use the 2,500-row CPU pilot and simulated event stream; replay is replacement
  batch application, not concurrent incremental throughput.}
  \label{fig:updates-staleness}
\end{figure*}

Figure~\ref{fig:updates-staleness}(a) times full-history replacement, not
sustained incremental ingestion.  Replaying 50, 100, 250, 500, and 740 events
took \ReplayFiftyMs{}, \ReplayOneHundredMs{}, \ReplayTwoFiftyMs{},
\ReplayFiveHundredMs{}, and \ReplaySevenFortyMs{}~ms, respectively, for a median
\MedianReplayRate{} and maximum \MaxReplayRate{} replayed events/s.  A full
re-embedding and rebuild took \RebuildMinMs{}--\RebuildMaxMs{}~ms.  The last
evaluated batch where replay was cheaper contained \ReplayLastFasterEvents{}
events, and the first where it was slower contained \ReplayFirstSlowerEvents{};
the present delta representation therefore
does not demonstrate update scalability.  A credible dynamic benchmark still
needs retained micro-batches, operation-stratified latency, compaction, and
concurrent query interference.

Panel (b) isolates watermark staleness on one 24-query cohort.  FGR-Exact,
given the complete history through each query day, retained at least
\StaleFGRMinRecall{} recall.
Immediately before each scheduled rebuild, the periodic view's recall fell
from \PeriodicDayThirtyFiveRecall{} at five days of lag to
\PeriodicDayFortyRecall{} at ten days and \PeriodicDayFortyFiveRecall{} at 15
days.  After the next rebuild it recovered to \PeriodicDayFiftyRecall{}, then declined to
\PeriodicDaySixtyRecall{} recall and \PeriodicDaySixtyPrecision{} precision at
day 60.  This supports the elementary point that an old watermark changes the
answer set; because event time and record age are simulated, it does not
quantify real-world source freshness.

\subsection{Diagnostic Stress Tests and Ablations}

\begin{figure*}[t]
  \centering
  \resizebox{\textwidth}{!}{\begin{tikzpicture}\definecolor{fgblue}{RGB}{0,114,178}\definecolor{fgorange}{RGB}{213,94,0}\definecolor{fggreen}{RGB}{0,158,115}\definecolor{fgpurple}{RGB}{204,121,167}
\begin{groupplot}[group style={group size=2 by 1,horizontal sep=1.1cm},width=0.47\textwidth,height=0.25\textwidth,grid=major,tick label style={font=\scriptsize},label style={font=\footnotesize},legend style={font=\scriptsize,draw=none}]
\nextgroupplot[xlabel={Replay day},ylabel={JSD},title={(a) Scripted drift trigger},legend pos=north west]
\addplot+[fgorange,mark=*] table[x=day,y=js_divergence,col sep=comma]{figures/data/drift_recovery.csv};\addlegendentry{JSD}
\addplot+[black,dashed,no marks] coordinates {(50,0) (50,0.6)};\addlegendentry{Rebuild}
\nextgroupplot[xlabel={Objects},ylabel={Serialized MB},title={(b) Index size},legend pos=north west]
\addplot+[fgblue,mark=*] table[x=objects,y=fgr_mb,col sep=comma]{figures/data/index_size_wide.csv};\addlegendentry{FGR}
\addplot+[fgorange,dashed,mark=square*] table[x=objects,y=graph_mb,col sep=comma]{figures/data/index_size_wide.csv};\addlegendentry{kNN graph}
\addplot+[fggreen,dotted,mark=triangle*] table[x=objects,y=spatial_mb,col sep=comma]{figures/data/index_size_wide.csv};\addlegendentry{Spatial}
\end{groupplot}\end{tikzpicture}}
  \caption{Scripted drift instrumentation and serialized footprint in the
  2,500-row CPU pilot with simulated replay.  The trigger is hard-coded and the
  graph line is a local exact-$k$NN proxy.}
  \label{fig:drift-size}
\end{figure*}
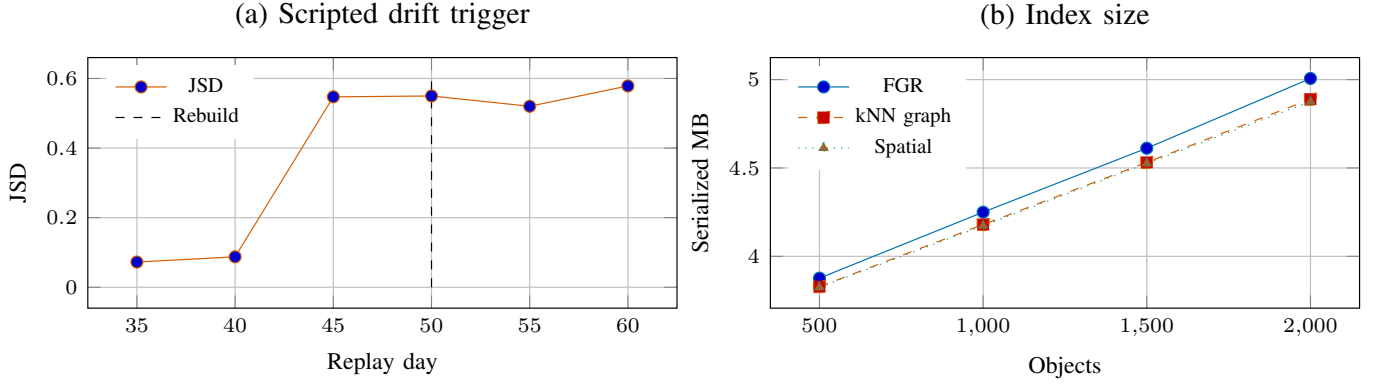

Figure~\ref{fig:drift-size}(a) is deliberately a scripted trigger test.  A
forced query-region shift raises Jensen--Shannon divergence from
\DriftDayFortyJSD{} on day 40 to \DriftDayFortyFiveJSD{} on day 45; a hard-coded
day-50 rebuild costs \DriftRebuildMs{}~ms.  The measured p95 then changes from
\DriftDayFortyFivePNinetyFiveMs{}~ms (day 45) to
\DriftDayFiftyPNinetyFiveMs{}~ms (day 50),
while the mean certificate changes from \DriftDayFortyFiveCert{} to
\DriftDayFiftyCert{}.  Because the workload changes and
there is no simultaneous no-rebuild counterfactual, this cannot be attributed
causally to recovery or used to validate an adaptive policy.  Panel (b) shows
serialized FGR size increasing from \SizeFGRFiveHundredMB{}~MB at 500 objects
to \SizeFGRTwoThousandMB{}~MB at 2,000, compared with
\SizeSpatialFiveHundredMB{} to \SizeSpatialTwoThousandMB{}~MB for spatial-first.
Four small points describe
the pilot footprint; they do not establish asymptotic scaling.

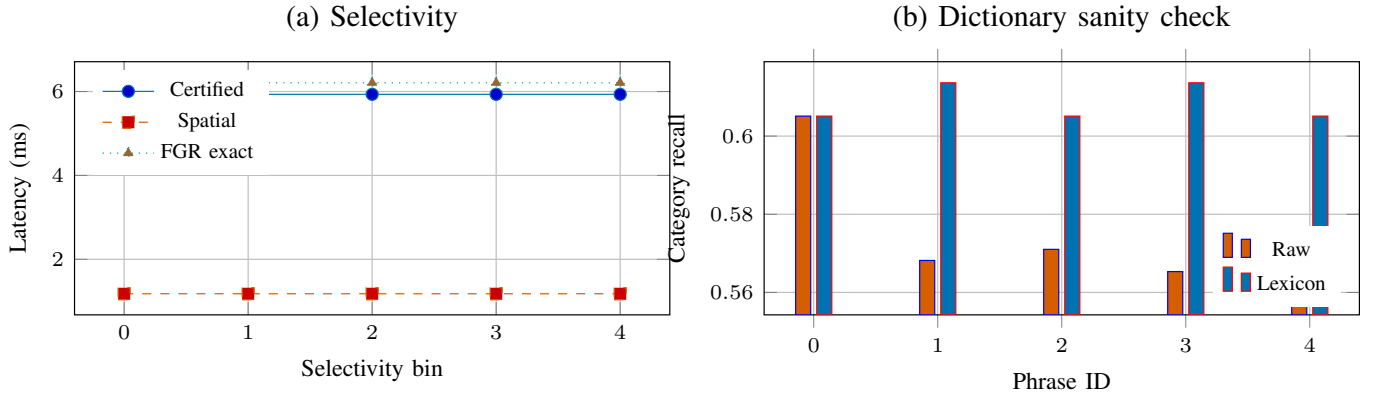
\begin{figure*}[t]
  \centering
  \resizebox{\textwidth}{!}{\begin{tikzpicture}\definecolor{fgblue}{RGB}{0,114,178}\definecolor{fgorange}{RGB}{213,94,0}\definecolor{fggreen}{RGB}{0,158,115}\definecolor{fgpurple}{RGB}{204,121,167}
\begin{groupplot}[group style={group size=2 by 1,horizontal sep=1.1cm},width=0.47\textwidth,height=0.25\textwidth,grid=major,tick label style={font=\scriptsize},label style={font=\footnotesize},legend style={font=\scriptsize,draw=none}]
\nextgroupplot[xlabel={Selectivity bin},ylabel={Latency (ms)},title={(a) Selectivity},legend pos=north west]
\addplot+[fgblue,mark=*] table[x=bin,y=fgr_cert95,col sep=comma]{figures/data/selectivity_wide.csv};\addlegendentry{Certified}
\addplot+[fgorange,dashed,mark=square*] table[x=bin,y=spatial_first_exact,col sep=comma]{figures/data/selectivity_wide.csv};\addlegendentry{Spatial}
\addplot+[fggreen,dotted,mark=triangle*] table[x=bin,y=fgr_exact,col sep=comma]{figures/data/selectivity_wide.csv};\addlegendentry{FGR exact}
\nextgroupplot[ybar,bar width=5pt,xlabel={Phrase ID},ylabel={Category recall},title={(b) Dictionary sanity check},legend pos=south east]
\addplot+[fill=fgorange] table[x=language_id,y=without_lexicon,col sep=comma]{figures/data/multilingual_wide.csv};\addlegendentry{Raw}
\addplot+[fill=fgblue] table[x=language_id,y=with_lexicon,col sep=comma]{figures/data/multilingual_wide.csv};\addlegendentry{Lexicon}
\end{groupplot}\end{tikzpicture}}
  \caption{Selectivity and dictionary-normalization diagnostics on the
  single-query airport pilot with simulated replay.  The lexical intervention
  is constructed and does not measure multilingual robustness.}
  \label{fig:selectivity-language}
\end{figure*}

Selectivity in Fig.~\ref{fig:selectivity-language}(a) exposes the same overhead:
FGR-Cert95 median latency rises from \SelectivityFGREmptyMs{}~ms for empty
answers to \SelectivityFGRHighMs{}~ms above 2\% selectivity, while spatial-first
remains between \SelectivitySpatialMinMs{} and \SelectivitySpatialMaxMs{}~ms.
FGR recall is at least \SelectivityFGRLowRecall{} through the 0.1--0.5\% bin and
\SelectivityFGRHighRecall{} above 2\%.  We do not
compare ``examined'' counts because their definitions differ by method.
Panel (b) is a dictionary-normalization sanity check, not a multilingual
retrieval evaluation.  The fixed lexicon raises category recall from
\FrenchRawRecall{} to \FrenchLexiconRecall{} for French,
\SpanishRawRecall{} to \SpanishLexiconRecall{} for Spanish,
\GermanRawRecall{} to \GermanLexiconRecall{} for German, and
\SynonymRawRecall{} to \SynonymLexiconRecall{} for an English synonym; English
remains \EnglishRawRecall{}.
Labels are circularly inferred from the word ``international,'' and tested
translations appear in the lexicon, so no language-generalization claim follows.

\begin{figure}[t]
  \centering
  \resizebox{\columnwidth}{!}{\begin{tikzpicture}\definecolor{fgblue}{RGB}{0,114,178}\definecolor{fgorange}{RGB}{213,94,0}\definecolor{fggreen}{RGB}{0,158,115}\definecolor{fgpurple}{RGB}{204,121,167}
\begin{groupplot}[group style={group size=1 by 1},width=0.92\columnwidth,height=0.34\columnwidth,grid=major,tick label style={font=\scriptsize},label style={font=\footnotesize},legend style={font=\scriptsize,draw=none}]
\nextgroupplot[ybar,bar width=5pt,xlabel={Variant ID},ylabel={Percent},legend pos=south west]
\addplot+[fill=fgblue,bar shift=-2.5pt] table[x=variant_id,y expr=100*\thisrow{recall},col sep=comma]{figures/data/ablation_plot.csv};\addlegendentry{Recall}
\addplot+[fill=fgorange,bar shift=2.5pt] table[x=variant_id,y expr=100*\thisrow{certificate},col sep=comma]{figures/data/ablation_plot.csv};\addlegendentry{Certificate}
\end{groupplot}\end{tikzpicture}}
  \caption{Guarantee ablation; variant IDs and latency are in
  Table~\ref{tab:ablation}. The stale-base variant is uncertified.}
  \label{fig:ablation}
\end{figure}
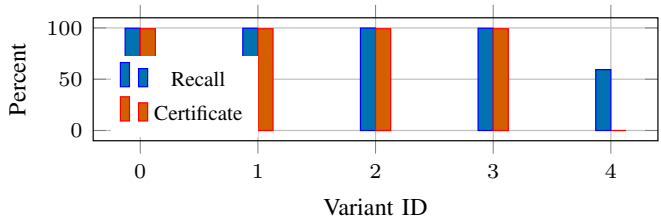

Finally, Table~\ref{tab:ablation} and Fig.~\ref{fig:ablation} show that removing
the graph proposer preserves recall (\FullAblationRecall{} versus
\NoGraphAblationRecall{}) and the mean certificate (\FullAblationCert{} versus
\NoGraphAblationCert{}) while changing latency from \FullAblationLatencyMs{} to
\NoGraphAblationLatencyMs{}~ms; the no-graph variant is faster in this run.
Removing semantic bounds or merging to one semantic block per spatial cell
changes latency to \NoSemanticAblationLatencyMs{} and
\OneBlockAblationLatencyMs{}~ms. These small, unreplicated ablations provide no
positive component-level speed evidence and cannot establish behavior at scale.
The decisive row is the deliberately unsafe stale-base variant: omitting the
delta lowers recall from \FullAblationRecall{} to \StaleNoDeltaRecall{}.  Its certificate is
suppressed because advancing the watermark without applying the events violates
the certificate premise.

\begin{table}[t]
\centering
\caption{Component ablation. The no-delta row is deliberately freshness-noncompliant, so no certificate is reported.}
\label{tab:ablation}
\small
\begin{tabularx}{\columnwidth}{@{}Xrrr@{}}
\toprule
Variant & Recall (\%) & Cert. (\%) & Latency (ms) \\
\midrule
Full & 99.68 & 99.20 & 8.17 \\
No graph proposer & 99.68 & 99.20 & 6.93 \\
No semantic bounds & 99.88 & 99.20 & 7.93 \\
One semantic block per cell & 99.68 & 99.20 & 8.18 \\
Stale base, no delta (uncertified) & 59.38 & -- & 4.73 \\
\bottomrule
\end{tabularx}
\end{table}

\subsection{Threats to Validity and Deployment Limits}

Local proxies, single timings, and Python code limit internal validity;
source-derived queries, simulated freshness, and serialized size limit
construct validity.  One uneven 2,500-airport sample, five seeds, and one CPU
limit external and statistical validity.  Single-writer replay omits
concurrency.  No personal data are used, but the pilot is not safety-ready.
Pinned artifacts aid reproduction.  Archived OSM
diffs~\cite{openstreetmap2026replication}, official baselines, independent
queries, concurrent updates, and broader property tests remain required.
Correctness passes, but efficiency fails: FGR-Cert95 is
\LatencyRatio$\times$ slower than spatial-first exact. Publication claims are
no-go.

\section{Conclusion}
\label{sec:conclusion}

FRESH-GEORANGE addresses semantic-spatial range enumeration when applications need every qualifying object from a current snapshot rather than an approximate ranking. The design combines geographic cells, semantic microblocks, admissible spatial and cosine bounds, a graph used for proposal ordering, and a latest-write delta overlay. Exact mode accounts for every nonprunable record; certified mode converts verified answers and an unresolved-record count into a deterministic recall lower bound. In the CPU pilot, exact mode returned the complete reference set on all \UniquePrimaryQueries{} queries, while the 95-percent mode reached \FGRCertRecall{} empirical mean recall with a \FGRCertBound{} reported mean certificate and no certificate violation. These correctness results are encouraging, but the efficiency evidence is negative: certified retrieval required \FGRCertMedianMs{} ms median latency, \LatencyRatio{} times the spatial-first exact baseline, and complete-history replay exceeded rebuild time for large batches. Consequently, this package demonstrates a functional proof of concept, not a submission-ready dynamic index or a deployment claim. Its limitation is scope: \PilotRawObjects{} real airport records were paired with simulated update histories and local proxy implementations rather than real OpenStreetMap diffs and official systems. Future work should implement incremental multi-version publication with sequence validation, then test million-scale diffs, concurrent workloads, independently judged semantic queries, and recall-matched official baselines.

\IEEEtriggeratref{17}
\bibliographystyle{IEEEtran}
\bibliography{references}

\end{document}